%% file: main.tex
\documentclass[3p]{elsarticle}
\usepackage[utf8]{inputenc}
\usepackage[T1]{fontenc}
\usepackage{amsmath}

\usepackage{circuitikz}
\usetikzlibrary{patterns}

\usepackage{hyperref}

\usepackage{subcaption}

\usepackage{mleftright}

\newcommand{\dif}{\mathrm{d}} 
\renewcommand{\i}{\mathrm{i}}
\newcommand{\e}{\mathrm{e}}
\definecolor{mplorange}{HTML}{ff7f0e}
\definecolor{mplblue}{HTML}{1f77b4}

\newcommand{\unit}[1]{\,\mathrm{#1}}

\journal{}
\begin{document}

\begin{frontmatter}

\title{Experimental observation of exceptional points in coupled pendulums}

\author[utc,isae-supmeca]{Nicolas Even}
\ead{nicolas.even@edu.supmeca.fr}
\author[isae-supmeca]{Benoit Nennig}
\ead{benoit.nennig@isae-supmeca.fr}
\author[utc]{Gautier Lefebvre}
\ead{gautier.lefebvre@utc.fr}
\author[utc]{Emmanuel Perrey-Debain}
\ead{emmanuel.perrey-debain@utc.fr}

\affiliation[utc]{organization={Universit\'e de technologie de Compi\`egne, Roberval (Mechanics, energy and electricity)},
addressline={Centre de recherche Royallieu},
postcode={CS 60319 - 60203},
city={Compiègne},
country={France}}

\affiliation[isae-supmeca]{organization={Laboratoire Quartz (EA7393), ISAE-Supméca – Institut supérieur de mécanique de Paris},
addressline={3 rue Fernand Hainaut},
postcode={93407},
city={Saint-Ouen},
country={France}}

\begin{abstract}

The concept of exceptional point (EP) is demonstrated experimentally in the case of a simple mechanical system consisting of two linearized coupled pendulums.
Exceptional points correspond to specific values of the system parameters that yield defective eigenvalues. These spectral singularities which are typical of non-Hermitian system means that both the eigenvalues and their associated eigenvectors coalesce. The existence of an EP requires an adequate parameterization of the dynamical system. For this aim, the experimental device has been designed with two controllable parameters which are the length of one pendulum and a viscous-like damping which is produced
via electromagnetic induction. Thanks to the observation of the free response 
of the coupled pendulums, most EP properties are experimentally investigated, showing good agreements with theoretical considerations. 
In contrast with many studies on EPs, mainly in the field of physics, the novelty of the present work is that controllable parameters are restricted to be real-valued,   and this requires the use of adequate search algorithms. 
Furthermore, it offers the possibility of exploiting the existence of EPs in time-domain dynamic problems.
\end{abstract}


\end{frontmatter}

\section{Introduction}
\label{sec:introduction}
Eigenvalues and eigenvectors are among the most used concepts for the analysis of linear  time-invariant systems.
In the context of mechanical vibrations, undamped systems, either discrete or continuous, are usually modelled by a set of equations which take the form of a mass and a stiffness matrix which have the remarkable property of being real-valued, symmetric and generally positive-definite. This matrix system belongs to the class of Hermitian pencils, which share most of the properties of the standard Hermitian eigenvalue problems. This means that eigenfrequencies are real-valued and the right eigenvectors, also called normal modes, define a mass orthogonal set of vectors \cite{lancaster2002}. 
Furthermore, algebraic and geometric multiplicity are equal, which means that the set of eigenvectors always define a basis. 
These  properties are the key ingredients for modal analysis in theoretical, computational and experimental vibration and acoustics. 

The treatment of damped systems is less trivial. In many situations, dissipation is modelled via an additional damping matrix, which can be handled by considering its effect as a perturbation of the lossless case or by assuming Rayleigh damping (also called proportional damping \cite{tisseur2001}). However, in order to address strong or more general dissipation mechanisms, and this also includes a class of viscoelastic damping models, approaches based on quadratic eigenvalue problems must be followed. 
In general, eigenfrequencies as well as eigenvectors become complex-valued as opposed to normal modes. An important point is that if all eigenvalues are distinct (or semisimple), eigenvectors still satisfy bi-orthogonality properties \cite{tisseur2001} and define a basis on which the solution of the problem, i.e. the free or the harmonic response,  can be expanded.
This can break down, however, if two (or more) eigenvalues coalesce. 

This fact is well documented in the study of non-Hermitian systems, arising notably in the field of quantum physics, optics and photonics \cite{moiseyev2011, heiss2012, miri2019}. 
One particularity of non-Hermitian matrices is the existence of singularities in the parameter space, known as \emph{exceptional points} (EPs) \cite{kato1966}, for which both eigenvalues and the associated eigenvectors coalesce at a branch point singularity, the latter becoming self orthogonal \cite{moiseyev2011}. Although the matrix cannot be diagonalized, spectral analysis can be carried out using Jordan decomposition, but additional generalized eigenvectors are required to compensate for the loss of one (or more) dimension(s) in the space spanned by the eigenvectors. 
In the vicinity of the EP, perturbation analysis shows that eigenvalues can be expanded as a Puiseux series~\cite{kato1966} showing a strong sensitivity of the eigenvalue with respect to the parameters.

For vibrating systems, frequency coalescence with complex parameter is known at least from the early '90s \cite{triantafyllou1991}, but the concept of EP, largely studied in the physics community, has yet not been fully explored in the field of applied mechanics though we can cite few pioneering works on the existence of multiple eigenvalues in non-conservative systems~\cite{seyranian1993}.
In recent papers on cavity acoustics, it has been found, both theoretically and experimentally, that EPs minimize the global time decay of the free response~\cite{kanev2018,bourquard2019,even2022}. These observations are in connection with the optimal design of tuned mass dampers (TMD) \cite{krenk2005,bisegna2012}.
The concept of optimal attenuation associated with EP also arise in duct acoustics since the pioneering work of Tester~\cite{tester1973} in 1973. 
More recent contribution and generalization can be found in~\cite{bi2015, xiong2017,perrey-debain2022,lawrie2022}. Here, eigenvalues are the axial wavenumbers of the duct acoustic modes and the attenuation along the duct is controlled by their imaginary part. When two (or more) duct acoustic modes coalesce for a particular choice of the complex-valued wall admittance(s), the imaginary part of the corresponding mode is found to reach a maximum.
The existence of EP is also linked to veering phenomena as shown in \cite{natsiavas1993} though the terminology does not yet appear explicitly in the paper. EPs are known to correspond to a branch point singularity in the complex plane, and this can have a negative impact on the radius of convergence of eigenvalue perturbation methods close to veering. To remedy this, an expansion around the EP using Puiseux series improves the convergence in the parameter space and this has been applied to the study of discrete random vibrating systems~\cite{ghienne2020}. More generally, a numerical algorithm has been devised in order to locate EP for linear or quadratic eigenvalue problems of large size~\cite{nennig2020} or depending on multiple real or complex parameters~\cite{mailybaev2006}.

A common property of EP is that it often corresponds to a boundary between dynamical regimes of the system. The simplest example is given by the classical damped harmonic oscillator, for which the critical damping is nothing else but an EP of the state-space matrix.
In the case of two coupled harmonic oscillators, the EP is the limit between weak and strong coupling \cite{seyranian2005,rodriguez2016,dolfo2018}. In the latter regime, the natural responses exhibit a beat, characteristics of periodic exchanges of energy between the two oscillators.
EPs are also encountered in stability analysis as discussed in the theoretical textbook~\cite{seyranian2003} with recent results in thermoacoustics~\cite{orchini2020}.
Finally, EP corresponds to the boundary between the $\mathcal{PT}$-broken and
unbroken phases in $\mathcal{PT}$ symmetric system \cite{bender2019}.

Recent experimental evidences of EP have been reported in many areas of physics: 
in microwave cavity \cite{dembowski2004}, in light–matter interactions in cavity magnon–polaritons \cite{zhang2017,zhang2019}, in chaotic optical microcavity \cite{lee2009}, in cryogenic optomechanical device \cite{xu2016}, 
with resistively coupled amplifying-LRC-resonator circuit \cite{choi2018}, with a mechanical resonator coupled with shunted piezoelectric patches \cite{li2023}, with electrically controlled interaction of light with organic molecules in the terahertz regime \cite{ergoktas2022} and in acoustics \cite{ding2016,shen2018}.
Mode coalescence of scattering matrices can also be observed, giving rise to unidirectional zero reflection~\cite{zhang2017,shen2018} and $\mathcal{PT}$-symmetry breaking transition~\cite{choi2018}. Many researchers explored the eigenvalue topological structure in the vicinity of EP via experimental Riemann surface reconstruction~\cite{lee2009,peng2014,xu2016,zhang2019a,ergoktas2022}.

To our knowledge, experimental observations of EPs in the field of applied vibration seems to be missing in the scientific literature.
The aim of this paper is to demonstrate  that it is possible to identify experimentally EPs in a simple linearized coupled pendulums system. Similar system has been concurrently used to investigate purely viscous coupling between oscillators \cite{lu2023}.
Here, the configuration has been devised in order to allow a direct observation  of most EP properties like the branch point structure in the parameter space, the coalescence of the eigenvalues and their corresponding eigenvectors, the specific form of the free response at EP and the energy exchange between the two oscillators. The existence of an EP relies on an appropriate parametrization of the dynamical system, here available parameters are either the inertia, the stiffness and/or the damping. 
The fact that these parameters are real-valued means that at least two controllable parameters are required. We show that we can reach the EP thanks to a fine-tuning of these parameters.
In the present work, the damping is controlled using eddy current brake mechanisms, whereas the restoring force is controlled by the length of one pendulum. The pendulums are weakly coupled and are almost identical, which means that the system is close to a critical point corresponding to a mode veering \cite{dieci2014,manconi2017} and this normally guarantees the existence of an EP in a certain neighborhood of the critical point \cite{wiersig2022}.

This paper is organized as follows.
First, the experimental set-up and the governing equations of the dynamical system are presented in Sec.~\ref{sec:problem_statement}.
Using modal analysis, a brief theoretical description of mode veering and mode coalescence
is given in Sec.~\ref{sec:modal_analysis_and_exceptional_points}. 
The dynamic modal identification technique is based on the analysis of our experimental data, i.e. the angular displacements of both pendulums, via appropriate fitting methods. Identification of the parameters of the model are carried out via optimization algorithms. All these procedures are explained in Sec.~\ref{sec:experimental_setup_and_calibration}.
Finally, experimental results showing most EP properties are presented and discussed in Sec.~\ref{sec:results}.

\section{Presentation of the experimental set-up and governing equations}
\label{sec:problem_statement}

\begin{figure}
    \begin{minipage}[c]{.5\linewidth}
        \centering
        \includegraphics[height=7cm]{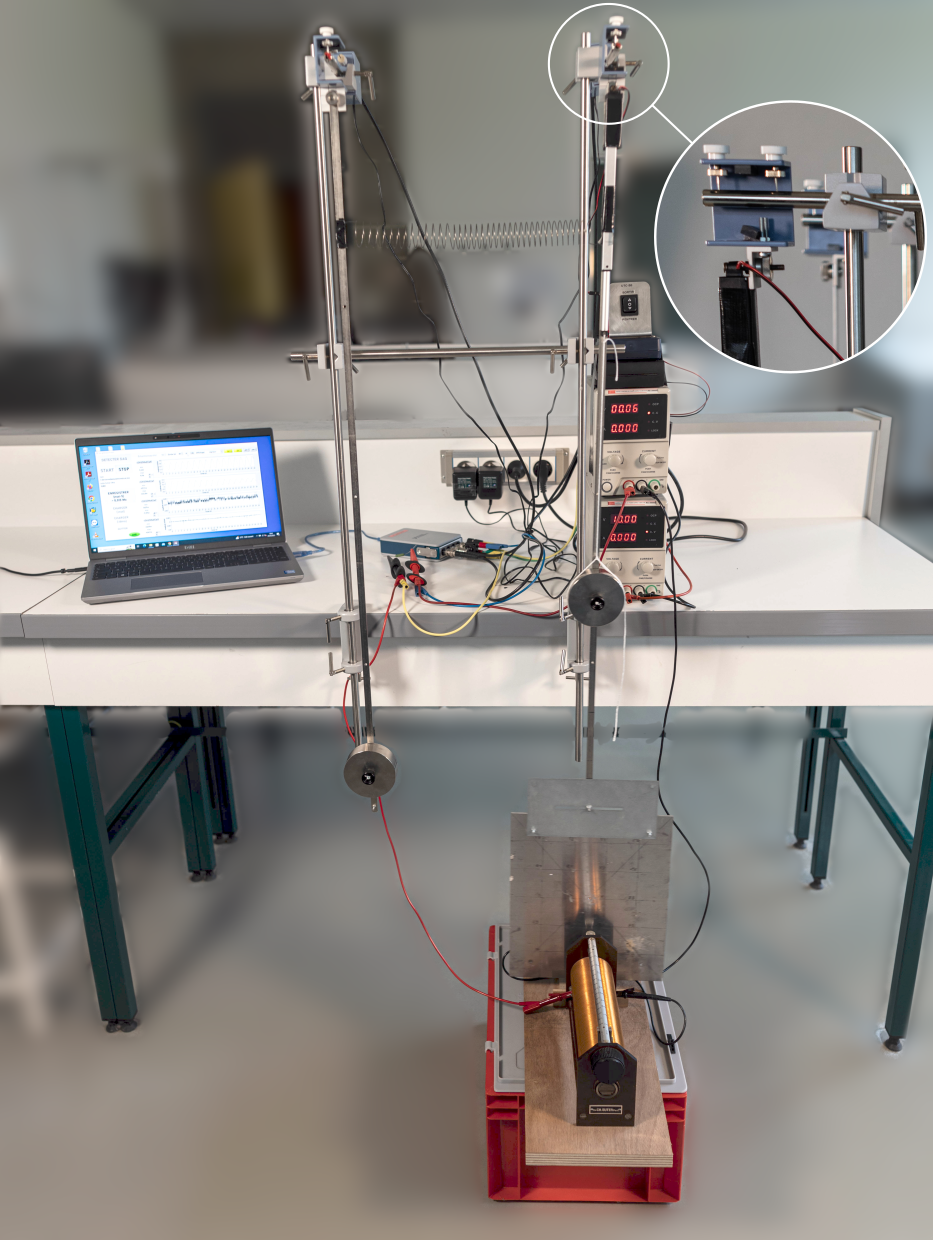}
        \subcaption{}
        \label{fig:experimental_setup}
    \end{minipage}%
    \begin{minipage}[c]{.5\linewidth}
        \centering
        \input{tikz/coupled_pendulums}
        \subcaption{}
        \label{fig:sketch}
    \end{minipage}
    \caption{(a) Experimental set-up and (b) sketch of the two coupled pendulums.}
    \label{fig:coupled_pendulums}
\end{figure}

Let us consider a system of two vertical pendulums coupled via a horizontal spring.
The experimental set-up for recording and evaluating the oscillations of the two  pendulums,  based on 3B Scientific (No.~8000563 UE1050600-230), is shown in Fig.~\ref{fig:experimental_setup} and its modeling in Fig.~\ref{fig:sketch}. Each pendulum consists of a disc-shaped weight of mass $m_i = 1\unit{kg}$ ($i=1,2$), that serves as the pendulum's bob, which can be moved along a flat rod to set the pendulum effective length $L_i$, before being fixed by means of a knurled screw. At the top of the flat rod is a U-shaped bracket, into which two needles are screwed to suspend the pendulum from a bearing rod with two grooves (see the inset). The two pendulums are coupled together by a spring of stiffness $k$ which is attached to the rods at a distance $d$ from the pivots. Thanks to a built-in Hall sensor in the bearing rod and a pair of magnets in the U-shaped bracket, an angle sensor detects the angle of deflection $\theta_i$ of the pendulum and converts it into a proportional electric signal sent to the NI-9223 voltage input module (No.~783284-01).

In order to perform a parametric analysis, the controllable parameters must be chosen so that the experimental set-up should be modified as little as possible while maintaining a high level of precision.
Thanks to the knurled screw, it is easy to fix $L_2$ at the desired length.
When it comes to viscous damping, eddy current damping is the one that is most suited to our set-up. To put this in place, a conductive plate is attached at the end of the second pendulum in order to create a magnetic damping by eddy currents induced by a magnetic field $\mathbf{B}$. The latter is generated by a pair of solenoids to control its magnitude by the electric current passing through them, varying between 0 and $0.5\,\mathrm{A}$. This damping device is therefore tunable, as it is proportional to the square of the current flowing through the coils, as described in \ref{app:magnetic_damping}.

As the inertia of the plate is not negligible, it must be taken into account in our calculations. From now, we will refer to the (full) pendulum as the combination of the bob and the compound pendulum, made up of the rod (and the plate for the second one) and characterized by $I_i$, $M_i$ and $R_i$, its moment of inertia about the pivot point, mass and distance between the pivot and its center of mass, respectively.
The differential equation which represents the motion of a single pendulum is

\begin{equation} \label{eq:pendulum}
    \mleft(m_i L_i^2 + I_i\mright) \ddot{\theta}_i + c_i \dot{\theta}_i + \mleft(m_i g L_i + M_i g R_i \mright) \sin  \theta_i  = 0,
\end{equation}
where $g$ is the standard acceleration due to gravity.
In the absence of magnetic field, damping coefficient $c_i$ accounts for residual loss due to air and mechanical friction.
Here we put
\begin{equation} \label{natural}
    \Omega_i = \sqrt{\frac{M_i g R_i}{I_i}}, \qquad \omega_{0i} = \sqrt{\frac{m_i g L_i + \Omega_i^2 I_i}{m_i L_i^2 + I_i}},
\end{equation}
the natural angular frequency for the compound pendulums and for full pendulums. Both pendulums are coupled together by the spring $k$ located at a distance $d$ from the pivots. The equations of motion take the form of the following system of linear ordinary differential equations (the small-angle approximation $\sin \theta_i  \approx \theta_i$ is used):
\begin{equation} \label{eq:system_of_odes}
    \begin{cases}
        \mleft(m_1 L_1^2 + I_1\mright) \ddot{\theta}_1 + (c_1 + c_{12}) \dot{\theta}_1 + \mleft(m_1 g L_1 + \Omega_1^2 I_1 + k d^2\mright) \theta_1 = c_{12} \dot{\theta}_2 + k d^2 \theta_2, \\
        \mleft(m_2 L_2^2 + I_2\mright) \ddot{\theta}_2 + (c_2 + c_{12}) \dot{\theta}_2 + \mleft(m_2 g L_2 + \Omega_2^2 I_2 + k d^2\mright) \theta_2 = c_{12} \dot{\theta}_1 + k d^2 \theta_1,
    \end{cases}
\end{equation}
where $c_{12}$ stems from a natural damping due to the coupling (considered negligible). The system can be conveniently recast in matrix form
\begin{equation} \label{eq:2nd-order_matrix_ode}
    \mathbf{M} \ddot{\mathbf{q}} + \mathbf{C} \dot{\mathbf{q}}  + \mathbf{K} \mathbf{q} = \mathbf{0},
\end{equation}
where $\mathbf{M}$, $\mathbf{C}$ and $\mathbf{K}$ stand for the mass, damping and stiffness matrices, respectively, and $\mathbf{q}$ is the generalized coordinates vector with
\begin{equation}\label{eq:system_matrices}
    \begin{gathered}
        \mathbf{M} = 
        \begin{bmatrix}
            m_1 L_1^2 + I_1 & 0 \\
            0             & m_2 L_2^2 + I_2
        \end{bmatrix},
        \qquad
        \mathbf{C} = 
        \begin{bmatrix}
            c_1 + c_{12} & -c_{12} \\
            -c_{12}      & c_2 + c_{12}
        \end{bmatrix},
        \\
        \mathbf{K} = 
        \begin{bmatrix}
            m_1 g L_1 + \Omega_1^2 I_1 + k d^2 & -k d^2 \\
            -k d^2 & m_2 g L_2 + \Omega_2^2 I_2 + k d^2
        \end{bmatrix},
        \qquad
        \mathbf{q} = 
        \begin{bmatrix}
            \theta_1 \\
            \theta_2
        \end{bmatrix}.
    \end{gathered}
\end{equation}
Eq.~\eqref{eq:2nd-order_matrix_ode} can be transformed into the equivalent first-order system using state-space representation
\begin{equation} \label{eq:1st-order_matrix_ode}
    \mathbf{A} \mathbf{p} = \mathbf{B} \dot{\mathbf{p}},
\end{equation}
using the first companion form \cite{tisseur2001}
\begin{equation}
    \mathbf{A} = 
    \begin{bmatrix}
        \mathbf{0}  & \mathbf{I} \\
        -\mathbf{K} & -\mathbf{C}
    \end{bmatrix},
    \quad
    \mathbf{B} = 
    \begin{bmatrix}
        \mathbf{I} & \mathbf{0} \\
        \mathbf{0} & \mathbf{M}
    \end{bmatrix},
    \quad
    \mathbf{p} = 
    \begin{bmatrix}
        \mathbf{q}\\
        \dot{\mathbf{q}}
    \end{bmatrix},
\end{equation}
suitable for modal analysis.

\section{Modal analysis, veering and exceptional points}
\label{sec:modal_analysis_and_exceptional_points}

\subsection{General theory}

The modal analysis of Eq.~\eqref{eq:2nd-order_matrix_ode} starts by considering the exponential ansatz $\mathbf{q} = \mathbf{v} \e^{\lambda t}$ which
yields the quadratic eigenvalue problem (QEP)
\begin{equation} \label{eq:qep}
    \mleft( \lambda^2 \mathbf{M} + \lambda \mathbf{C}  + \mathbf{K} \mright) \mathbf{v} = \mathbf{0},
\end{equation}
where $\mathbf{v}$ is an eigenvector associated with the eigenvalue $\lambda$.  Alternatively, the modal analysis can be done by solving the generalized eigenvalue problem (GEP)
\begin{equation} \label{eq:gep}
    \mathbf{A} \mathbf{x} = \lambda \mathbf{B} \mathbf{x} \quad \text{with }
    \mathbf{x} =
    \begin{bmatrix}
        \mathbf{v}\\
        \lambda \mathbf{v}
    \end{bmatrix}.
\end{equation}
By construction, eigenvalues are the roots of the characteristic polynomial
\begin{equation} \label{eq:characteristic_polynomial}
    p(\lambda) = \det \mleft(\lambda^2 \mathbf{M} + \lambda \mathbf{C}  + \mathbf{K}\mright) = \det (\mathbf{A} - \lambda \mathbf{B}).
\end{equation}
Because matrices $\mathbf{M}$, $\mathbf{C}$ and $\mathbf{K}$ are real, eigenvalues and eigenvectors are real or come in complex conjugate pairs \cite{tisseur2001}. Using the Laplace transform notation, we may write the complex eigenvalue as $\lambda = -\sigma \pm \i \omega$, where $\omega$ corresponds to the natural angular frequency and $\sigma$ to the exponential decay constant, both being real. 
When all roots are simple, which is generally the case, the free response of the system can be expressed as the linear combination of modes
\begin{equation} \label{eq:mode_superposition_not_ep}
    \mathbf{q} (t) = a_1 \mathbf{v}_1 \e^{\lambda_1 t}
    + a_2 \mathbf{v}_2 \e^{\lambda_2 t}
    + \mathrm{c.c.} = 2 \operatorname{Re} \mleft (a_1 \mathbf{v}_1 \e^{\lambda_1 t}
    + a_2 \mathbf{v}_2 \e^{\lambda_2 t} \mright ),
\end{equation}
where coefficients $a_1$ and $a_2$ are complex-valued and `c.c.' means complex conjugate. In this paper, we are interested in situations where an eigenvalue becomes a double root, which signifies
\begin{equation} \label{doubleroot}
    p(\lambda) = \partial_\lambda p (\lambda) = 0.
\end{equation}
In this case, the nature of the solution depends on whether the eigenvalue is semisimple or defective (see Tisseur \cite{tisseur2001}). A semisimple eigenvalue means that there are two distinct eigenvectors associated with the double root. 
This generally corresponds to degeneracies from conservative systems, also called \emph{diabolical points} \cite{seyranian2005}, and are known to arise in symmetric configurations.
Eigenvalues are defective when the geometric multiplicity does not match the algebraic multiplicity, and their existence requires that the system is non-Hermitian due to some dissipation (or gain) mechanism.

\subsection{Undamped system}

To illustrate this, consider first the undamped case with $\mathbf{C}= \mathbf{0}$, eigenvalues have the closed form expression
\begin{equation}\label{eq:undamped_lda}
    -\lambda^2_i = \omega^2_i = \frac{X_1 + X_2
    +(-1)^i \sqrt{(X_1 - X_2)^2+4\kappa_1\kappa_2} }{2},
\end{equation}
where 
\begin{equation}
\kappa_i = \frac{kd^2}{m_i L_i^2 + I_i},
\end{equation}
stands for the coupling strength and $X_i = \omega^2_{0i} + \kappa_i$. Because all coefficients involved are all real-valued and positive, it is clear that the occurrence of a double root is possible only if the system is uncoupled $kd^2=0$ with 
\begin{equation}
    \lambda_\mathrm{c} =\i \omega_{c} = \i \omega_{01} = \i \omega_{02},
\end{equation}
which correspond to a critical point\footnote{This is borrowed from the work of Mace \cite{manconi2017}.} (for clarity, subscript `c' will refer to such a situation). Here, eigenvectors are simply given by $\mathbf{v}_1 =[1,0]^\mathrm{T}$ and $\mathbf{v}_2 =[0,1]^\mathrm{T}$ and 
\begin{equation} 
    \mathbf{q}_\mathrm{c} (t) = 2 \operatorname{Re} \mleft ((a_1 \mathbf{v}_1 
    + a_2 \mathbf{v}_2) \e^{\lambda_\mathrm{c} t} \mright ).
\end{equation}

When the system is weakly coupled, i.e. $kd^2$ is sufficiently small, the first resonant frequency is not modified $\omega_1 = \omega_{01}$ and $\omega_2 = \omega_{01} + \mathcal{O}(kd^2)$. Perturbations of parameters in the vicinity of the critical point is a typical mode veering in weakly coupled systems, as discussed in \cite{manconi2017}. Experimental results shown in Sec.~\ref{sec:results}
will serve to illustrate this scenario.

\subsection{Damped system}\label{sec:damped_th}

When damping is present, the analysis of the problem is less trivial, as closed form expressions are generally not available. However, the behavior around the double root can still be described. Let us accept for the moment that there exist some specific values of the controlling parameters $(\mu, \nu)$ which leads to the existence of a double root Eq.~\eqref{doubleroot}. By expanding the determinant as a Taylor series, it can be shown that (see~\ref{app:taylor_series}) : 
\begin{equation} \label{Puiseux}
\lambda_i \approx \lambda_0 + (-1)^{i} \sqrt{\Delta},
\end{equation}
where 
\begin{equation} \label{Delta}
\Delta = -2 \frac{\partial_{\mu} p (\lambda_0,\mu_0, \nu_0) (\mu -  \mu_0) +  \partial_{\nu} p (\lambda_0,\mu_0, \nu_0) (\nu -  \nu_0)}{\partial^2_{\lambda} p (\lambda_0,\mu_0, \nu_0)},
\end{equation}
is a small quantity.
The pair $(\mu_0, \nu_0)$ defines an Exceptional Point (EP) and for clarity, subscript ‘0’ will refer to this situation. The square root behavior in Eq.~\eqref{Puiseux} means that eigenvalues in the vicinity of an EP are extremely sensitive, in contrast with the regular behavior around the critical point\footnote{Although there is a square root in Eq.~\eqref{eq:undamped_lda}, its argument is always positive as opposed to Eq.~\eqref{Puiseux}.}. Furthermore, an EP is associated with a defective eigenvalue which means that the free response contains an additional  term growing linearly in $t$ and in fact
\begin{equation} \label{eq:mode_superposition_ep}
    \mathbf{q}_0 (t) =  2 \operatorname{Re} \mleft ((a_0 \mathbf{v}_0 + a_1 (\tilde{\mathbf{v}}_0+ \mathbf{v}_0 t )) \e^{\lambda_0 t} \mright ) .
\end{equation}
The growing term, which is a characteristic of the critical damping of the harmonic oscillator, can be understood from a physical point of view since it corresponds to  the beating phenomena in the limit case where both eigenfrequencies are equal. This term also appears formally as the matrix exponential of the operator, written in its Jordan form~\cite{bossart2021}.
Here,  $\mathbf{v}_0$ is the unique eigenvector (up to a norm) satisfying
\begin{equation}
    \mleft( \lambda_0^2 \mathbf{M} + \lambda_0 \mathbf{C}  + \mathbf{K} \mright) \mathbf{v}_0 = \mathbf{0},
\end{equation}
and $\tilde{\mathbf{v}}_0$ is a generalized eigenvector of rank 2, which verifies (see \cite[p.~251]{tisseur2001}):
\begin{equation} \label{eq:gen_egv}
    \mleft( \lambda_0^2 \mathbf{M} + \lambda_0 \mathbf{C}  + \mathbf{K} \mright) \tilde{\mathbf{v}}_0 = - \mleft(2 \lambda_0 \mathbf{M} + \mathbf{C} \mright) \mathbf{v}_0.
\end{equation}
Note the Jordan chain $\{\mathbf{v}_0, \tilde{\mathbf{v}}_0\}$ given by Eq.~\eqref{eq:gen_egv} is not unique, as $\{a \mathbf{v}_0, a \tilde{\mathbf{v}}_0 + b \mathbf{v}_0\}$  is also a Jordan chain. The uniqueness of the Jordan chain is usually ensured by imposing the normalization $\|\mathbf{v}_0\| = 1$ as well as the orthogonality condition $\tilde{\mathbf{v}}_0^{\mathrm{H}} \mathbf{v}_0 = 0$ which is performed via a Gram–Schmidt process.

The form of the solution given in Eq.~\eqref{eq:mode_superposition_ep} means  that if the initial conditions are collinear with $\mathbf{v}_0$ then $a_1=0$, there is no growing term and the solution remains collinear with $\mathbf{v}_0$. If $a_1 \neq 0$,  vector $\mathbf{v}_0$ will still dominate the long-time response because
\begin{equation} \label{eq:mode_superposition_ep_approx}
    \mathbf{q}_0 (t) \approx  2 \operatorname{Re} \mleft ( a_1  \mathbf{v}_0 t  \, \e^{\lambda_0 t} \mright ) \quad \mathrm{as } \  t \rightarrow  \infty.
\end{equation}
This means that the dynamical systems acts as a polarizer~\cite{bossart2021}, where any initial state converts into the final state $\mathbf{v}_0$.
The theoretical analysis of double roots (semisimple and defective) in simple dynamical mechanical systems such as the one presented here has been the subject of many papers in the scientific literature \cite{triantafyllou1991, ryu2015, dolfo2018, seyranian2003, ghienne2020}. While EPs are relatively easy to construct by letting parameters of the model, such as a geometrical length \cite{triantafyllou1991} or a stiffness coefficient \cite{dolfo2018, ghienne2020}, to be complex-valued, the novelty of this work lies in the existence of EPs with real-valued parameters, here the length and the coefficient of the magnetic damping. 
However, finding the specific values at the EP, i.e. $L_2=L_0$ and $c_2=c_0$, assuming that they exist, is not a trivial task. In some cases which involves only two degrees of freedom as in~\cite{ryu2015}, closed form solutions are sometimes available. This is not the case here, although approximate expressions, which are relevant for the present work, are provided in \ref{app:identical_pendulums} for the specific case of two identical pendulums.
A general search algorithm able to treat large algebraic systems have been proposed recently by some of the present authors in~\cite{nennig2020} and the interested reader is referred to the open source software \href{https://github.com/nennigb/EasterEig}{\texttt{eastereig}} \cite{eastereig}. The technique relies on the analyticity of certain functions in the complex plane, which allow to obtain EP with complex-valued parameters.
In the present paper, we propose a variant approach given in \ref{app:real-valued_exceptional_point} in order to find double roots with two real-valued controllable parameters.

\section{Calibration, data extraction and location of the EP in the parameter space}
\label{sec:experimental_setup_and_calibration}

In order to locate an EP, accurate value for all parameters of the model Eq.~\eqref{eq:system_of_odes} are needed. This section presents the different identification steps using uncoupled and coupled pendulum and global modal (eigenvalues and eigenvectors) extraction from time series. 

\subsection{Uncoupled system}

\begin{figure}[ht]
    \centering
    \includegraphics{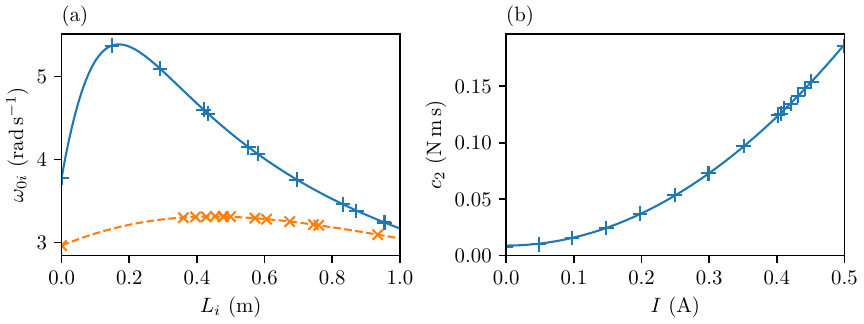}
    \caption{(color online) (a) Natural angular frequencies $\omega_{0i}$ for each pendulum as a function of $L_i$. The blue solid line and the orange dashed line are obtained by fitting Eq.~\eqref{natural} to data points corresponding to the first (blue \textcolor{mplblue}{$+$}) and second pendulum (orange \textcolor{mplorange}{$\times$}), respectively. (b) Rotational damping of the second pendulum as a function of the current, the blue solid line is obtained by fitting Eq.~\eqref{eq:damping_per_current} to data points (blue \textcolor{mplblue}{$+$}).}
    \label{fig:uncoupled_pendulums}
\end{figure}

The first step is to identify parameters associated with each pendulum, treated separately and uncoupled. In the under-damped case, the solution has the exact form
\begin{equation} \label{eq:damped_sine_wave}
    \theta_i (t) = A_i \e^{-\zeta_i \omega_{0i} t} \sin \mleft(\sqrt{1 - \zeta_i^2} \omega_{0i} t + \varphi_i \mright),
\end{equation}
where $\omega_{0i}$ is the natural frequency from Eq.~\eqref{natural}. The amplitude $A_i$ and the phase $\varphi_i$  depend on  the initial conditions and the damping ratio  $\zeta_i$ is given by :
\begin{equation} \label{eq:natural_angular_frequency_and_damping_ratio}
    \zeta_i =\frac{c_i}{2\sqrt{(m_i L_i^2 + I_i)  (m_i g L_i + \Omega_i^2 I_i)}}.
\end{equation}
Using curve-fitting algorithms with non-linear least squares (NLS) method, the four optimal values of the parameters involved in Eq.~\eqref{eq:damped_sine_wave} are returned as well as their covariance matrix. First, the bob is removed from the pendulum, which gives directly $\Omega_i = \omega_{0i}$. A series of measurements is then conducted with different values of $L_i$ (the latter is measured with a ruler with less than 1\,mm of error). From the results, given in Fig.~\ref{fig:uncoupled_pendulums}a, the moment of inertia $I_i$ is recovered.

Identified parameters are reported in Tab.~\ref{tab:uncoupled_pendulums_parameters} (values are given within a 95\% confidence interval using $\pm$ twice the standard deviation). As expected, the moment of inertia for the first pendulum is much smaller than the second one.
\begin{table}[ht]
    \caption{Parameters identified by curve fitting and their 95\% confidence intervals.}
    \centering
    \begin{tabular}{c c c c}
        \hline
        $I_1$ & $I_2$ & $\Omega_1$ & $\Omega_2$ \\
        ($\mathrm{kg}\,\mathrm{m}^{2}$) & ($\mathrm{kg}\,\mathrm{m}^{2}$) &  ($\mathrm{rad}\,\mathrm{s}^{-1}$) & ($\mathrm{rad}\,\mathrm{s}^{-1}$) \\
        \hline
        $0.056271 \pm 0.000249$ & $1.000150 \pm 0.004381$ & $3.782120 \pm 0.00000$5 & $2.963880 \pm 0.000012$ \\
        \hline
    \end{tabular}
    \label{tab:uncoupled_pendulums_parameters}
\end{table}
In the presence of magnetic damping, another series of measurements is conducted by increasing the current $I$ flowing through the coils.
The measured damping coefficients are indicated in Fig.~\ref{fig:uncoupled_pendulums}b
and the fact that it grows quadratically goes along with the theory, as explained in \ref{app:magnetic_damping}. Using the same NLS method to fit
\begin{equation} \label{eq:damping_per_current}
    c_2 (I) = a I^2 + b,
\end{equation}
we get the values for the positive constant $a$ and the residual damping $b$ useful to initialize the system.

\subsection{Coupled system}\label{sec:coupled_identification}

In a second step, the same curve fitting algorithm as before is applied to the response of the coupled system from Eq.~\eqref{eq:mode_superposition_not_ep},
\begin{equation*}
    \mathbf{q} (t) = 2 \operatorname{Re} \mleft (a_1 \mathbf{v}_1 \e^{\lambda_1 t}
    + a_2 \mathbf{v}_2 \e^{\lambda_2 t} \mright ),
\end{equation*}
where coefficients $a_{i}$ are complex-valued and depend on the initial conditions.
Although other  high-resolution methods for complex exponential recovering, like LS-ESPRIT \cite{roy1986}, could have been used, it was observed that fitting methods exploiting the full signal length, performed slightly better here.

\begin{figure}[ht]
    \centering
    \includegraphics{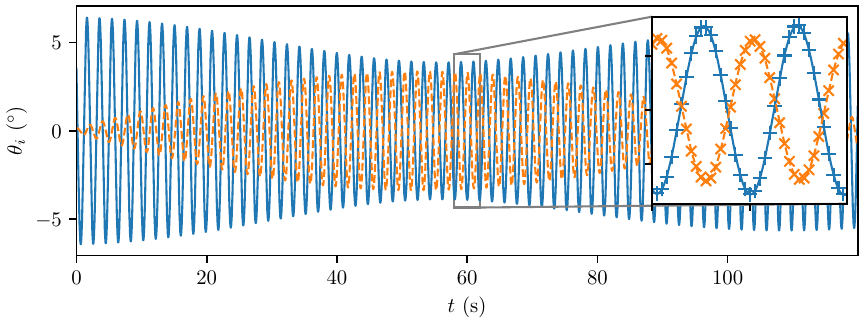}
    \caption{Evolution of the angular position of the first (blue \textcolor{mplblue}{$+$}) and second pendulum (orange \textcolor{mplorange}{$\times$}). The blue solid line and the orange dashed line are obtained by fitting Eq.~\eqref{eq:mode_superposition_not_ep} to data points of the first and second pendulum, respectively. The identified eigenvalues are $\lambda_1 = -0.002354 \pm 0.000023 - (3.224085 \pm 0.000023)\i$ and $\lambda_2 = -0.000933 \pm 0.000009 - (3.280612 \pm 0.000009)\i$ with a coefficient of determination $R^2 = 0.99967$. For the sake of readability, only one data point out of five is shown.}
    \label{fig:free_response}
\end{figure}

An example is given in Fig.~\ref{fig:free_response} for a 2-minute-long signal sampled at $50\,\mathrm{Hz}$ and obtained after moving only the first pendulum from its equilibrium position. 
This excitation allows to populate all the modes and will be used in all the following experiments. The experimental signal and the modal fit are nearly identical.

\noindent Note that the method allows one to identify the eigenvalues and their associated non-normalized eigenvectors $\hat{\mathbf{v}}_i = a_i\mathbf{v}_i$.  For small damping, it was observed that the standard deviation is of order $10^{-5}$ for both eigenvalues and eigenvectors and can reach $10^{-4}$ for highly damped signals.

After identification, we introduce the matrices
\begin{equation} \label{eq:XD}
    \hat{\mathbf{X}}  = \begin{bmatrix}
        \hat{\mathbf{v}}_1  & \hat{\mathbf{v}}_2 & \hat{\mathbf{v}}^*_1  & \hat{\mathbf{v}}^*_2\\
        \hat{\lambda}_1 \hat{\mathbf{v}}_1 &  \hat{\lambda}_2 \hat{\mathbf{v}}_2 &
        \hat{\lambda}^*_1 \hat{\mathbf{v}}^*_1 &  \hat{\lambda}^*_2 \hat{\mathbf{v}}^*_2
    \end{bmatrix},
    \qquad
 \hat{\mathbf{\Lambda}}  = \mathrm{diag}( \hat{\lambda}_1, \hat{\lambda}_2,\hat{\lambda}^*_1, \hat{\lambda}^*_2),
\end{equation}
where hat symbol signifies that these are measured quantities.
Parameters of the coupled system are linked to these quantities via the equality 
\begin{equation} \label{eq:eigendecomposition}
    \mathbf{B}^{-1} \mathbf{A} = \mathbf{X} \mathbf{\Lambda} \mathbf{X}^{-1},
\end{equation}
where hat symbol is omitted as it corresponds to the exact eigendecomposition of the idealized model~\eqref{eq:system_of_odes}.
Parameters appear explicitly, since
\begin{equation} \label{BA}
    \mathbf{B}^{-1} \mathbf{A} =
    \begin{bmatrix}
        0 & 0 & 1 & 0 \\
        0 & 0 & 0 & 1 \\
        -\frac{m_1  g L_1 + \Omega_1^2 I_1 + k d^2}{m_1 L_1^2 + I_1} & \frac{k d^2}{m_1  L_1^2 + I_1} & -\frac{c_1 + c_{12}}{m_1  L_1^2 + I_1} & \frac{c_{12}}{m_1  L_1^2 + I_1} \\
        \frac{k d^2}{m_2 L_2^2 + I_2} & -\frac{m_2 g L_2 + \Omega_2^2 I_2 + k d^2}{m_2 L_2^2 + I_2} & \frac{c_{12}}{m_2 L_2^2 + I_2} & -\frac{c_2 + c_{12}}{m_2 L_2^2 + I_2}
    \end{bmatrix}.
\end{equation}
The optimization problem consists in minimizing the cost function
\begin{equation}\label{eq:optim}
    {\underset {\{L_1, L_2, k d^2,c_1, c_2 , c_{12}\} }{\operatorname {arg\,min} }}\; \lVert \mathbf{B}^{-1} \mathbf{A}  - \hat{\mathbf{X}}\hat{\mathbf{\Lambda}} \hat{\mathbf{X}}^{-1} \rVert_{\mathrm{F}},
\end{equation}
using the Sequential Least Squares Programming (SLSQP) method. Here, $\lVert \cdot  \rVert_{\mathrm{F}}$ denotes the Frobenius norm, 
and  $\{L_1, L_2, k d^2,c_1, c_2 , c_{12}\}$ is the most general set of the unknown parameters.
In order to estimate the confidence interval of the different parameters, a Quasi-Monte Carlo method (with 30000 samples) is used 
to propagate the uncertainty of the eigenvalues $\hat{\mathbf{\Lambda}}$ and the eigenvectors $\hat{\mathbf{X}}$ with the covariance matrix given by the NLS algorithm.
Thus, the evaluated confidence intervals characterize the robustness of the dynamic model, represented by the matrix \eqref{BA}, and of the optimization process \eqref{eq:optim}.
Constant parameters are identified by setting $L_2\approx 0.7\,\mathrm{m}$ in absence of magnetic damping (note this value roughly corresponds to the critical point $L_c$ and its associated EP $L_0$ as shown later). 
The optimization is performed, and their mean values are listed in Tab.~\ref{tab:coupled_pendulums_parameters}.
The residual damping coefficient $c_2(0)$ of the second pendulum, though nearly negligible, is higher by an order of magnitude than $c_1$ and this is attributed to the fluid resistance due to the movement of the conductive plate. The damping coefficient between the two oscillators $c_{12}$ is the smallest one and can be ignored.

\begin{table}[ht]
    \caption{Parameters identified by minimization and their 95\% confidence intervals.}
    \centering
    \begin{tabular}{c c c c c}
        \hline
        $L_1$ & $k d^2$ & $c_1$ & $c_2(0)$ & $c_{12}$ \\
        ($\mathrm{m}$) & ($\mathrm{N}\,\mathrm{m}$) & ($\mathrm{N}\,\mathrm{m}\,\mathrm{s}$) & ($\mathrm{N}\,\mathrm{m}\,\mathrm{s}$) & ($\mathrm{N}\,\mathrm{m}\,\mathrm{s}$) \\
        \hline
        $0.953412 \pm 0.000014$ & $0.175062 \pm 0.000081$ & $\sim 10^{-3}$ & $\sim 10^{-2}$ & $\sim 10^{-4}$ \\
        \hline
    \end{tabular}
    \label{tab:coupled_pendulums_parameters}
\end{table}

At this point, we should point out that the series of  measurements presented in the next section are performed by first setting the pendulum length $L_2$ and then by increasing the damping coefficient $c_2(I)$. In all cases, all constant parameters are taken from Tab.~\ref{tab:coupled_pendulums_parameters}, and the two controllable parameters $L_2$ and $c_2$ are obtained from the optimization problem Eq.~\eqref{eq:optim}.
It was observed that for small damping, i.e. without magnetic damping, the 95\% confidence intervals are of order of $10^{-5}$ for both parameters whereas the precision slightly deteriorates by an order of magnitude for highly damped signals and this is in line with what was observed earlier for both eigenvalues and eigenvectors.

\subsection{Location of the EP in the parameter space}\label{sec:ep_location}

The existence of an EP in the parameter space relies on the roots of the polynomial system  $\mathbf{f} (\lambda; c_2, L_2)$ (see Eq.~\eqref{eq:polsys}). In order to find an EP with real-valued parameters, a search algorithm developed in \ref{app:real-valued_exceptional_point} is used. Results, computed with fixed parameters identified earlier, are given in Tab.~\ref{tab:eps_locations1}. We note that two EPs have been found, and this could have been anticipated since the dynamical system admits two critical points. This is discussed in the next section.
\begin{table}[ht]
    \caption{Exceptional points (labelled with subscript ‘0’) computed with the search algorithm, the double root is defined as $\lambda_0 = -\sigma_0 \pm \i \omega_0$.}
    \centering
    \begin{tabular}{c c c c c}
        \hline
     &  $\omega_0$ & $\sigma_0$ & $L_0$ & $c_0$ \\
     & ($\mathrm{rad}\,\mathrm{s}^{-1}$) & ($\mathrm{s}^{-1}$) & ($\mathrm{m}$) & ($\mathrm{N}\,\mathrm{m}\,\mathrm{s}$) \\
        \hline
    EP$_1$ &    3.271158  & 0.027053  & 0.241573 & 0.113217  \\
    EP$_2$ &    3.271194  & 0.023155  & 0.674831 & 0.133028  \\
        \hline
    \end{tabular}
    \label{tab:eps_locations1}
\end{table}

\section{Results}\label{sec:results}

In this section, the behavior of the dynamical system is explored by varying the two controllable parameters $L_2$ and the magnetic damping $c_2(I)$. The scenario without magnetic damping is first investigated.

\subsection{Lossless system}

\begin{figure}[ht]
    \centering
    \includegraphics{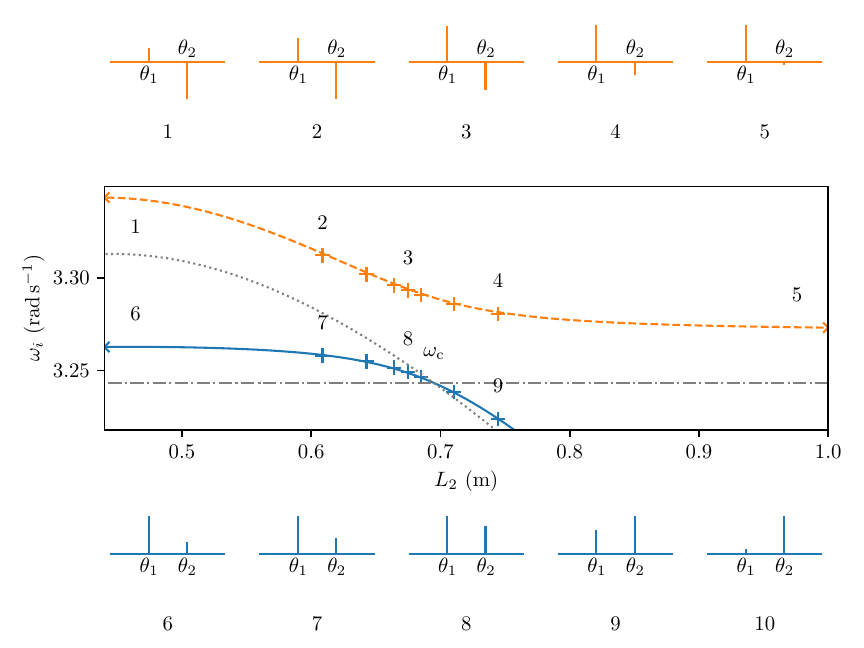}
    \caption{(color online) Eigenvalues of the system and their associated eigenvectors as a function of $L_2$ in the undamped case. The blue solid line (resp. orange dashed line) corresponds to the in-phase (resp. out-of-phase) mode with their associated eigenvectors below (resp. above). The $+$ markers correspond to the eigenvalues identified from the experiment, while the $\times$ markers correspond to the numerical data points.}
    \label{fig:veering_xp}
\end{figure}

Here, the system is not damped, or at most marginally, thus eigenvectors of the QEP are excepted to be real-valued and nearly orthogonal.
Measured eigenfrequencies $\omega_i$ and eigenvectors of the coupled system as a function of $L_2$ are displayed  in Fig.~\ref{fig:veering_xp}. 
The analytical natural frequency of the second uncoupled pendulum $\omega_{02}(L_2)$ is also shown (gray dotted line) as well as $\omega_{01}$ (gray dash-dotted line) which is a constant. The phenomenon of avoided crossing \cite{triantafyllou1991} can be observed in the vicinity of the critical points.
The location of the latter is found by solving $\omega_{01}=\omega_{02}(L_c)=\omega_{\mathrm{c}}$ (given in Eq.~\eqref{natural}) which yields the quadratic equation
 \begin{equation} 
\omega^2_{\mathrm{c}} m_2 L_{\mathrm{c}}^2 - m_2 g L_{\mathrm{c}} =   (\Omega_2^2 - \omega^2_{\mathrm{c}})I_2,
\end{equation}
giving two distinct values for the length of the second pendulum, see Tab.~\ref{tab:critical_locations}. These values are very close to those associated with the 2 EPs identified in Tab.~\ref{tab:eps_locations1}.
These results reflect a typical mode veering in weakly coupled systems which was expected since the normalized coupling strengths, i.e.
$\kappa_i/\omega_{0i}^2$, do not exceed $2\%$. The corresponding eigenvectors, i.e. the in-phase and out-of-phase modes identified by the color, are also illustrated. Two observations can be made: first, the two pendulums tend to be uncoupled when the length of the second pendulum deviates notably from its critical value $L_c$ and second, modes rotate rapidly around the point where veering occurs. The exchange of the eigenvectors components near a  critical point is well illustrated in Fig.~\ref{fig:veering_xp} where the fast oscillating mode dominated by the second pendulum (labelled 1) becomes dominated by the first pendulum (labelled 5) as $L_2$ increases.
Detailed discussions on this topic can be found in~\cite[p.~184]{dieci2014} or in \cite{manconi2017} for instance.

\begin{table}[ht]
    \caption{Critical points corresponding to the uncoupled system. These values are calculated with $L_1 =0.953 \,\mathrm{m}$.}
    \centering
    \begin{tabular}{c c c}
        \hline
        $\omega_\mathrm{c}$& \multicolumn{2}{c}{$L_2=L_{\mathrm{c}}$} \\
        ($\mathrm{rad}\,\mathrm{s}^{-1}$) & \multicolumn{2}{c}{($\mathrm{m}$)}\\
        \hline
        3.243473 & 0.237523 & 0.694657 \\
        \hline
    \end{tabular}
    \label{tab:critical_locations}
\end{table}

\subsection{Dynamics of the damped coupled pendulums}

\begin{figure}[ht]
    \centering
    \includegraphics{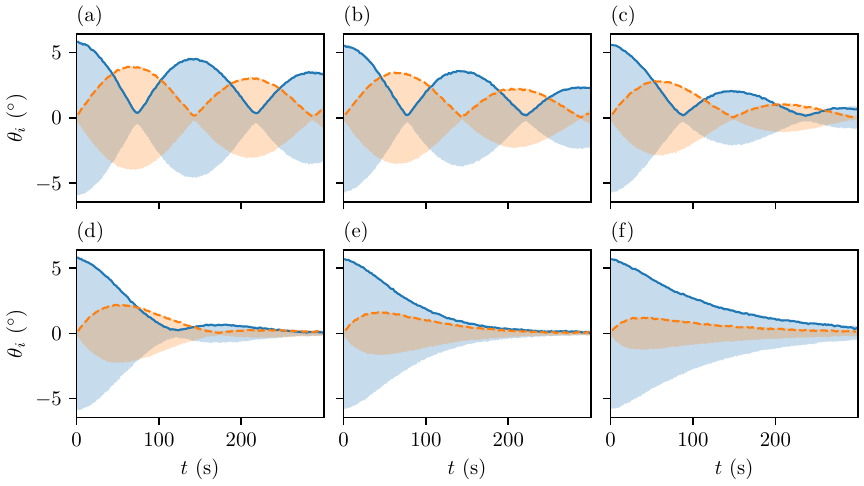}
    \caption{(color online) Experimental free response envelopes of the first (blue solid line) and second pendulum (orange dashed line) after moving only the first pendulum from its equilibrium position for different current intensities varying from 0 to $0.5\,\mathrm{A}$. (a) $\alpha_2=0.06$, (b) $\alpha_2=0.12$, (c) $\alpha_2=0.28$, (d) $\alpha_2=0.54$, (e) $\alpha_2=0.93$ and (f) $\alpha_2=1.35$ where $\alpha_2 = c_2 / c_0$.}
    \label{fig:beat}
\end{figure}

In Fig.~\ref{fig:beat} are shown the time evolution of the two pendulums for various values of the magnetic damping coefficient, whilst the value of the other controllable parameter is kept constant and $L_2=0.6739\,\mathrm{m}$ which nearly corresponds to the theoretical value of the EP$_2$ in Tab.~\ref{tab:eps_locations1}. This value is selected because it is associated with the smallest exponential decay and this is better suited to analyze the signals. All results correspond to the free-response, whereby the first pendulum is initially displaced from its equilibrium position without initial velocity. The second pendulum starts to swing due to the coupling as expected.
The envelope of each signal, computed via a Hilbert transform, is also shown for better visualization of the energy exchanges. We identify three different coupling regimes: (i) when $c_2$ is below $c_0$ (Fig.~\ref{fig:beat}a-d) the pendulums transfer energy between them faster than they each dissipate energy, thus displaying beatings with an energy exchange period proportional to $(\omega_2-\omega_1)^{-1}$; (ii) in the vicinity of the EP (Fig.~\ref{fig:beat}e), the energy exchange disappears due to the mode coalescence showing very strong attenuation rate and this case is shown (see \ref{app:identical_pendulums}) to correspond to a balance between the damping ratio and the normalized coupling strength; (iii) for very large values of the magnetic damping (Fig.~\ref{fig:beat}f), the second pendulum tends to stay still while the first one oscillates with small attenuation.
Regime (i) is sometimes called `strong coupling'  in contrast with the `weak coupling' regime of scenario (iii) and the EP acts as a boundary between these two regimes~\cite{rodriguez2016,dolfo2018}. Note that the existence of EP is not a necessary condition, as similar observations can be made in the situation where both pendulums are identical and when both damping coefficients are equal \cite{rodriguez2016}. In this scenario, the transition between the two regimes is observed when $c_2 \approx c_0/2$.

\begin{figure}[ht]
    \centering
    \includegraphics{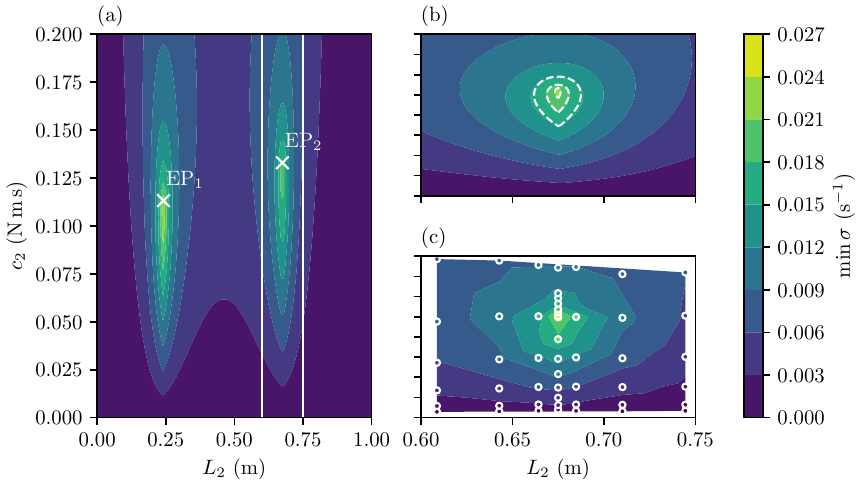}
    \caption{(color online) Exponential decay constants of the least attenuated mode in the parameter space. (a) Numerical results from the model with the EPs locations (b) close-up view with isolines (white dashed line) obtained by the Puiseux series~\eqref{eq:nuclear_core} (c) close-up view  showing the experimental results (the surface plot is interpolated from experimental data, indicated by the markers).}
    \label{fig:min_sigma}
\end{figure}

EPs are often linked to a maximal value of attenuation. Here, we can directly check this fact by evaluating the exponential decay of the least attenuated mode for a large set of values of the controllable parameters. Experimental data reported in Fig.~\ref{fig:min_sigma}c show a clear maximum at the EP. This is confirmed numerically in Fig.~\ref{fig:min_sigma}a
showing the existence of two EPs, both associated with the two critical points (see Tab.~\ref{tab:critical_locations}).
In the vicinity of the second EP a closer view is shown in Fig.~\ref{fig:min_sigma}b. We obtain a good agreement with Eq.~\eqref{eq:nuclear_core} given in \ref{app:taylor_series}, based on a development of the eigenvalues around the EP. This allows to identify the domain of validity of the local approximation (the square root term in Eq.~\eqref{eq:nuclear_core} is only valid in a very close proximity to the EP and linear terms are required). We note in passing that although the optimal global attenuation is demonstrated for the free-response of the system, 
a strong connection with tuned mass dampers (TMD) is expected. In this regard, one can cite~\cite{bisegna2012} or~\cite{krenk2005} where it is shown that the double root condition represents an upper limit for the damper. 
For the configuration treated in~\cite{krenk2005}, their optimal solution or the one obtained using the standard Den Hartog's approach is found when $c_2 \approx c_0/2$ which means that the existence and the location of an EP in the parameter space should provide some useful information for the design of the TMD in more complex situations.

\begin{figure}[ht]
    \centering
    \includegraphics{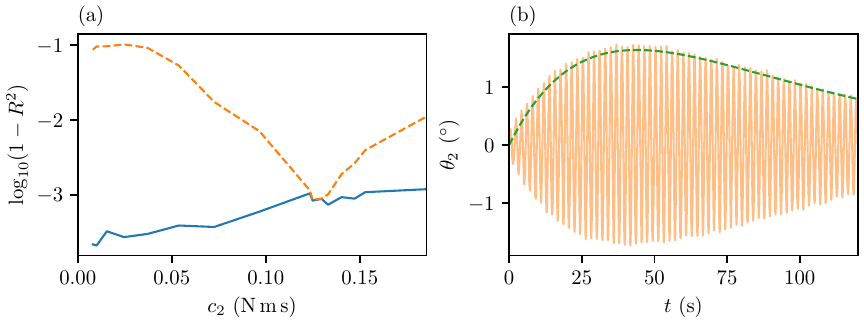}
    \caption{(color online) (a) Coefficient of determination as a function of $c_2$ for the fits of Eq.~\eqref{eq:mode_superposition_not_ep} (blue solid line) and Eq.~\eqref{eq:mode_superposition_ep} (orange dashed line). (b) Free response of the second pendulum near the EP after moving only the first pendulum from its equilibrium position (orange solid line), as well as the envelope $a_1 v_{0,2} t \e^{\lambda_0 t} + \mathrm{c.c.}$ (green dashed line).}
    \label{fig:response_ep_jordan}
\end{figure}

By following the theoretical considerations given in  Sec.~\ref{sec:damped_th}, the solution is expected to be of a different nature at EP with a growing term in $t$, see Eq.~\eqref{eq:mode_superposition_ep}. 
In this regard, it is instructive to apply the NLS algorithm for the classical solution Eq.~\eqref{eq:mode_superposition_not_ep}  and \eqref{eq:mode_superposition_ep}
The coefficient of determination $R^2$ is shown in Fig.~\ref{fig:response_ep_jordan}a.  Clearly, the fit with the Jordan chain of Eq.~\eqref{eq:mode_superposition_ep}
is nearly perfect in a close vicinity of the EP and degrades quickly otherwise.
This special solution can be identified visually in the specific case where $\theta_2(0) = 0$ because we have simply:
\begin{equation}
    \theta_2(t) =  2 \operatorname{Re} \mleft (a_1 v_{0,2} t \e^{\lambda_0 t} \mright) = 2 |a_1 v_{0,2}| \sin(\omega_0 t + \varphi_2) \, t \, \e^{-\sigma_0 t},
\end{equation}
where $v_{0,2}$ is the second component of the eigenvector $\mathbf{v}_0$,
which allows a direct observation, see dotted line in Fig.~\ref{fig:response_ep_jordan}b, of the growing term and the deviation from the pure exponential behavior (in fact, the pure exponential decay can be observed only if the initial conditions are collinear with $\mathbf{v}_0$).

\subsection{Exploring the topology of the eigenvalues}\label{subsec:topology}

\begin{figure}[ht]
    \centering
    \includegraphics{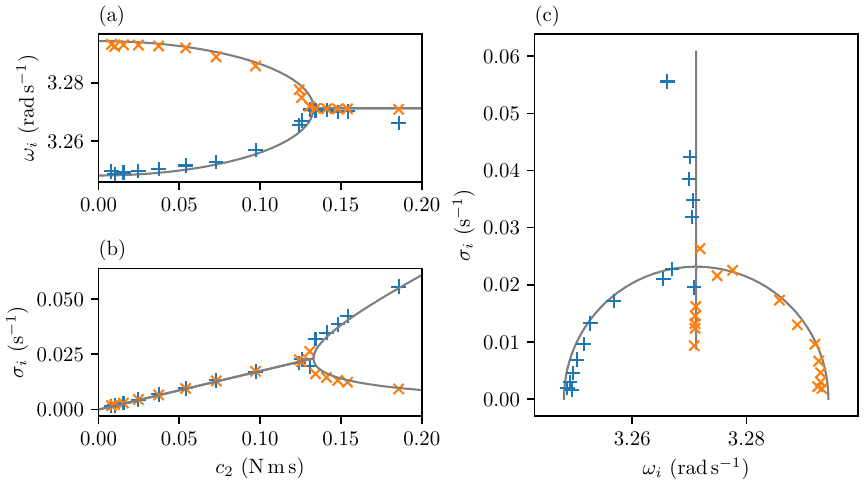}
    \caption{(color online) (a) Natural angular frequencies and (b) exponential decay constants of the eigenvalues as a function of $c_2(I)$ and (c) parametric representation in the complex eigenvalue plane. The markers are the measured data points corresponding to the first (blue \textcolor{mplblue}{$+$}) and second mode (orange \textcolor{mplorange}{$\times$}) while the gray solid line correspond to Eq.~\eqref{eq:eigenvalues_approx}.}
    \label{fig:eigenvalues_loci}
\end{figure}

In this first series of experiments, the length $L_2$ is set to $0.6750\,\mathrm{m}$ which corresponds to EP$_2$. The magnetic field is then gradually increased and for a selected set of values of the  damping coefficient $c_2$, a modal analysis is performed following the procedure described in the previous section. 
For each experiment, both measured and computed eigenvalues are shown in Fig.~\ref{fig:eigenvalues_loci}a-b.
Note that computed values are obtained by performing an eigendecomposition of the linear system~\eqref{eq:gep}.
Experimental data, indicated with markers, are shown to be in very good agreement with the numerical evaluation (in straight lines). In the absence of magnetic damping, the exponential decay constant is very small as expected, and natural angular frequencies are distinct although very close, their ratio being $\omega_2/\omega_1 \approx 1.013$. 
When the damping coefficient increases, the exponential decay constants, which are nearly identical, grow linearly until eigenvalues coalesce when $c_2 \approx c_0$, which was expected.
Above this value, exponential decay constants split whereas angular frequencies remain identical. In Fig.~\ref{fig:eigenvalues_loci}c, the same results are shown in  the $\omega\,\sigma$\nobreakdash-plane (which is equivalent to show the path of the two eigenvalues in the complex-plane). The pattern is typical of a branch point singularity of the square root well-known for EPs, which appears in the Puiseux series~\eqref{Puiseux}.
This indicates that eigenvalues are 
infinitely sensitive as $c_2$ tends to $c_0$, and this fact has been exploited for sensing applications like in \cite{wiersig2022}. That this should provide a maximum of attenuation, at least locally, can be understood from a graphical point of view as the two eigenvalues coalesce in opposite directions as confirmed in Fig.~\ref{fig:eigenvalues_loci}c.

We can derive an approximate solution for the eigenvalues around the EP, by exploiting the fact that both pendulums have similar characteristics and are weakly coupled. In this regard, the interested reader is referred to the theoretical developments given in  \ref{app:identical_pendulums}. It emerges that the eigenvalue at EP is approximately given by
\begin{equation}
\lambda_0 \approx -\sigma_0 + \i \omega_0 =-\omega_{\mathrm{c}}\epsilon/2 
 + \i \omega_{\mathrm{c}}(1+ \epsilon/2).
\end{equation}
where $\epsilon = (\kappa_1+\kappa_2)/(2 \omega^2_{\mathrm{c}})$ is the normalized coupling strength and 
\begin{equation}
c_0 \approx 2 (m_2 L_c^2 +I_2) \omega_{\mathrm{c}} \epsilon.
\end{equation}
The comparison with exact values calculated with the search algorithm (see Tab.~\ref{tab:eps_locations1}) shows that these first order approximations are very reliable as long as $\epsilon$ does not exceed a few percent. This is confirmed in Tab.~\ref{tab:eps_locations} where  approximated values are reported. Note that the exact value $L_0$ deviates slightly from the critical point $L_{\mathrm{c}}$ because the ideal condition that $\kappa_1=\kappa_2$, assumed in \ref{app:identical_pendulums}, is not satisfied here. 

\begin{table}[ht]
    \caption{Exceptional points calculated with the perturbation theory developed in \ref{app:identical_pendulums}.}
    \centering
    \begin{tabular}{c c c c c}
        \hline
     &  $\omega_0$ & $\sigma_0$ & $L_c$ & $c_0$ \\
     & ($\mathrm{rad}\,\mathrm{s}^{-1}$) & ($\mathrm{s}^{-1}$) & ($\mathrm{m}$) & ($\mathrm{N}\,\mathrm{m}\,\mathrm{s}$) \\
        \hline
    EP$_1$ &    3.270226 &  0.026752  & 0.237523  & 0.113063 \\
    EP$_2$ &    3.266555 &  0.023082  & 0.694657 & 0.136892 \\
        \hline
    \end{tabular}
    \label{tab:eps_locations}
\end{table}

The trajectories of the eigenvalues with respect to $c_2$ can also be predicted by the theory:
\begin{equation} \label{eq:eigenvalues_approx}
\lambda_i \approx \i \omega_0 -\sigma_0 \left (\frac{c_2}{c_0} + (-1)^i \sqrt{\left (\frac{c_2}{c_0}\right)^2-1} \right ). 
\end{equation}
This closed-form expression~\eqref{eq:eigenvalues_approx} for the eigenvalues, even if the two pendulums slightly differ in our case, provides an excellent estimation, compared with the direct computation from the eigendecomposition. The two computations do not differ in Fig.~\ref{fig:eigenvalues_loci}.

\begin{figure}[ht]
    \centering
    \includegraphics{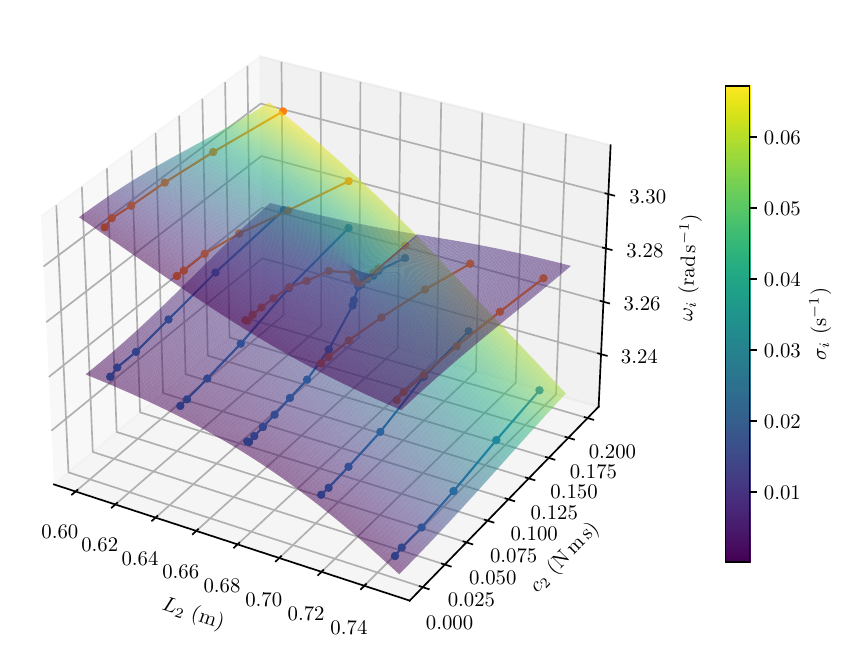}
    \caption{(color online) Riemann surface of the complex eigenvalues in the parameter space. The elevation corresponds to the angular frequency and the color to the exponential decay, thus the EP occurs when Riemann surfaces intersect and when both surfaces have the same color. The surface plots are obtained using the numerical model and experimental data correspond to the markers.}
    \label{fig:riemann_surface}
\end{figure}

Now, in order to get the full picture, we display the two complex eigenvalues as a function of the two controllable parameters $L_2$ and $c_2$. For this purpose, it is convenient to use a Riemann surface representation where the natural angular frequencies are plotted on the vertical axis and the value of the exponential decay constants is indicated by the color map on each surface, this is shown in Fig.~\ref{fig:riemann_surface}. Note that the surface plots are calculated using numerical eigenvalue solver and experimental data are indicated by the markers and that both results agree very well even in the vicinity of the EP, where eigenvalues are known to be more sensitive. Again these results show that the EP acts as a square branch point singularity in the $(L_2,c_2)$ plane, although this statement is not exact from a mathematical point of view since a function of two real parameters is not equivalent to a function of a simple complex variable\footnote{Here eigenvalues cannot be viewed as function of the complex-valued variable $L_2 + \i c_2$ and our calculations show that
$$\Delta \approx 
-(-0.000580782-0.0109833 \i) \left [ L_2 -  L_0
+ (-0.0370022+1\i) (\bar{c}_2 -  \bar{c}_0) \right ]$$
where the bar symbol signifies that the damping coefficient has been conveniently scaled and $\bar{c}_2 = 0.704579 c_2$.
This equation means that, in a close vicinity of the EP, eigenvalues are \emph{almost} a function of $L_2 + \i \bar{c}_2$. When dealing with complex-valued parameters, the analysis is simplified since the square root singularity is exact. To see this, it suffices to consider a \emph{single}
parameter, $L_2$ for instance, and extend its value in the complex plane with $L_2=\mathrm{Re}(L_2) + \i \mathrm{Im}(L_2)$ (although this is unrealistic, complex-valued lengths are considered in~\cite{triantafyllou1991})}.

\begin{figure}[ht]
    \centering
    \includegraphics{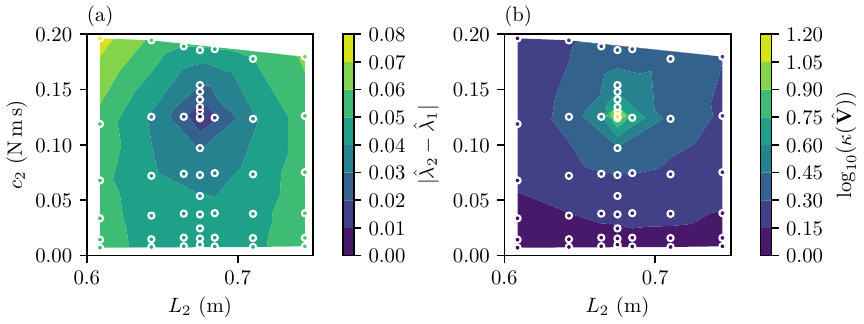}
    \caption{(color online) Distance between the measured eigenvalues (a) and condition number of the experimental modal matrix $\hat{\mathbf{V}}$ (b) with respect to $L_2$ and $c_2$. Surface plots are interpolated from experimental data, indicated by the markers.}
    \label{fig:condiff}
\end{figure}

Exceptional points also correspond to the coalescence of the eigenvectors as it can be shown that (see \cite[p.~36]{seyranian2003})
\begin{equation} \label{Puiseuxvec}
\mathbf{v}_i\approx \mathbf{v}_0+  (-1)^{i} \sqrt{\Delta} \, \tilde{\mathbf{v}}_0.
\end{equation}
A convenient way to quantify the departure from orthogonality \cite{{trefethen2005}} due to the damped nature of the system, is to consider the condition number of the modal matrix
\begin{equation} \label{cond}
\kappa(\hat{\mathbf{V}}) = \| \hat{\mathbf{V}}\|_2 \| \hat{\mathbf{V}}^{-1}\|_2 \quad \textrm{with} \ \hat{\mathbf{V}} = 
\begin{bmatrix}
\hat{\mathbf{v}}_1  & \hat{\mathbf{v}}_2    
\end{bmatrix},
\end{equation}
where it is reminded that hat symbol signifies that these are measured quantities of Sec.~\ref{sec:coupled_identification}. 
This indicator is well adapted to our experimental data, as it provides a robust and useful tool in order to identify the location of the EP. 
We want to emphasize that this characterization does not require any prior modelling of the dynamical system, and can be viewed as a purely experimental signature of the EP.
The condition number of the modal matrix also measures  the degree of non-normality of a matrix \cite[Chap. 48]{trefethen2005} with strong link to the sensitivity of the eigenvalues due to perturbations \cite[Bauer-Fike theorem (2.19)]{trefethen2005}. By construction, the condition number should be equal to 1 (this holds if  vectors have the same amplitude) in the lossless scenario, whereas it should tend to infinity at the EP. To be more precise, the conditioning number is expected to behave like $\kappa = \mathcal{O}(|\Delta|^{-1/2})$.  Results are shown in Fig.~\ref{fig:condiff}b. In Fig.~\ref{fig:condiff}a are also indicated the distance between the measured eigenvalues $|\hat{\lambda}_1 - \hat{\lambda}_2|$, which behave like $|2\sqrt{\Delta}|$ near the EP. 

\section{Conclusion}\label{sec:conclusion}
The concept of exceptional point (EP) has been demonstrated experimentally in the case of a simple vibration system consisting of two linearized coupled pendulums.
The two pendulums are chosen to be nearly identical and weakly coupled, which means that the system is close to a critical point corresponding to a mode veering, and this normally guarantees the existence of an EP. The latter is detected via an appropriate parameterization of the dynamical system. For this aim, the experimental device has been designed with two controllable parameters which are the length of one pendulum and a viscous-like damping which is produced via electromagnetic induction.
Experimental data, which are the angular displacements, are processed using curve fitting algorithms and this allows one to identify the eigenfrequencies  and their associated eigenvectors.
A series of measurements is undertaken in the parameter space by varying the two controllable parameters.
The two complex eigenfrequencies which must be interpreted as a function of these parameters are displayed using a Riemann surface representation.
The coalescence of eigenvectors is detected via the conditioning number of the two vectors, which, in theory, becomes infinite at EP. In a close vicinity of the EP, the two eigenvalues behave like a Puiseux series, i.e. with fractional exponents, and experimental results are shown to be in good agreement with theoretical predictions. This means that the dynamical system is very sensitive with respect to the parameters.
Three interesting properties which are often associated with exceptional points have been observed: first, an EP corresponds to a boundary between the weak and the strong coupling regime which exhibits periodic exchanges of energy between the two oscillators. Second, EPs minimize the global time decay of the free response and therefore provide an optimal dissipation. Third, EPs act as a polarizer, since the unique eigenvector dominates the long-time response.

What distinguishes the present work from most studies is the fact that controllable parameters are real-valued. This has important consequences for at least two reasons. From a practical point of view, it allows us to envisage other type of vibration systems encountered in real-life applications. From a theoretical point of view, the singularity of the EP in the parameter space is not a branch point of a complex square root and there are situations for which EPs belong to a continuous curve in the parameter space. The best example, described in Appendix, is the symmetric configuration whereby both pendulums are identical and weakly coupled. 

We hope that the research work presented here has highlighted the interdisciplinary significance of exceptional points, and in particular, their importance in mechanical vibration. The strong link with optimal attenuation in dissipative dynamical systems is also clearly shown. Though the analysis presented here is restricted to the free response, there are good reasons to believe that the concept could be used in order to minimize the forced response of a vibrating structure. In this regard, the connection with tuned mass dampers remains to be explored.

\appendix

\section{Eddy current brake}
\label{app:magnetic_damping}

We consider a conductive plate of volume $V$ rotating at an angular velocity $\dot{\theta}$ around an axis $\mathbf{e}_{z}$ and subjected to a magnetic field $\mathbf{B} = B \mathbf{e}_{z}$. At a distance $r$ from the axis of rotation, the induced eddy currents are given by Ohm's law
\begin{equation}
    \mathbf{j} = \kappa \mathbf{v} \times \mathbf{B} = \kappa r \dot{\theta} B \mathbf{e}_{r},
\end{equation}
where $\mathbf{j}$, $\kappa$ and $\mathbf{v} = r \dot{\theta} \mathbf{e}_{\theta}$ are the current density, electrical conductivity and the instantaneous velocity. The Lorentz force density acting on the plate is therefore
\begin{equation}\label{eq:Lorentz}
    \mathbf{f} = \mathbf{j} \times \mathbf{B}  = - \kappa r \dot{\theta} B^{2} \mathbf{e}_{\theta},
\end{equation}
and the corresponding torque reads
\begin{equation}
    \boldsymbol{\tau} = \int_{V} r \mathbf{e}_r \times \mathbf{f} \, \dif V = - \kappa B^{2} \dot{\theta} \int_{V} r^{2} \mathbf{e}_z \, \dif V.
\end{equation}
Thus, the damping torque is proportional to the angular velocity and is also proportional to the square of the magnetic flux density. For a solenoid of length $l$ with $N$ turns with a ferromagnetic core of effective permeability $\mu_{\mathrm{eff}}$, the magnetic flux is directly proportional to the current $I$:
\begin{equation}
    B = \mu_0 \mu_{\mathrm{eff}} \frac{N I}{l},
\end{equation}
where $\mu_0$ is the vacuum magnetic permeability.

\section{Approximate solution in the vicinity of an exceptional point}
\label{app:taylor_series}
The starting point is to consider the Taylor expansion of the determinant with respect to the eigenvalue $\lambda$ and two controllable parameters which, for the sake of generality, we shall call $\mu$ and $\nu$.  In the vicinity of an exceptional point $(\lambda_0,\mu_0,\nu_0)$, we can write formally
\begin{equation}
p(\lambda,\mu, \nu) = \sum_{i=0}^\infty \sum_{j=0}^\infty \sum_{k=0}^\infty
\frac{p^0_{\lambda^i \mu^j \nu^k}}{i! j! k!} (\delta \lambda)^i 
(\delta\mu)^j  (\delta \nu)^k,
\end{equation}
where we introduce
\begin{equation}
p^0_{\lambda^i \mu^j \nu^k} = \partial^i_\lambda  
\partial^j_\mu \partial^k_\nu p(\lambda_0,\mu_0,\nu_0),
\end{equation}
and 
\begin{equation}
\delta \lambda =\lambda-\lambda_0, \quad  \delta \mu = \mu-\mu_0 \quad \textrm{and} \quad \delta \nu =\nu-\nu_0.
\end{equation}
By construction, $p(\lambda,\mu, \nu)$ is zero everywhere, thus the double root condition leads to the equation
\begin{multline} \label{eq:second-order_taylor_expansion}
 p^0_\mu \delta \mu + p^0_\nu \delta \nu 
    + \frac{1}{2} \mleft[ p^0_{\lambda^2}(\delta \lambda)^2 + p^0_{\mu^2} (\delta \mu)^2 + p^0_{\nu^2} (\delta \nu)^2\mright] \\
    +  p^0_{\lambda \mu} \delta \lambda \delta \mu +  p^0_{\lambda \nu} \delta \lambda \delta \nu +   p^0_{\mu \nu} \delta \mu \delta \nu  + \frac{1}{6} p^0_{\lambda^3}(\delta \lambda)^3  + \cdots =0.
\end{multline}

By following earlier work \cite{perrey-debain2022}, an approximate closed-form solution can be found by searching the roots of the polynomial in the form
\begin{equation}
    \delta \lambda = \pm \mleft(- 2 \frac{p_\mu^0 \delta \mu + p_\nu^0 \delta \nu}{p^0_{\lambda^2}}\mright)^{1/2} + a \delta \mu + b \delta \nu,
    \label{eq:nuclear_core}
\end{equation}
where coefficients $a$ and $b$ remain to be found. This is easily done by injecting this expression in Eq.~\eqref{eq:second-order_taylor_expansion} and one obtains

\begin{equation}
    a = -\frac{p_{\lambda \mu}^0}{p_{\lambda^2}^0}
   + \frac{1}{3} \frac{p_{\lambda^3}^0 p_\mu^0}{(p^0_{\lambda^2})^2},
\end{equation}
and
\begin{equation}
    b = -\frac{p_{\lambda \nu}^0}{p_{\lambda^2}^0}
   + \frac{1}{3} \frac{p_{\lambda^3}^0 p_\nu^0}{(p^0_{\lambda^2})^2}.
\end{equation}

\section{Existence of exceptional points for two identical pendulums weakly coupled}\label{app:identical_pendulums}

\subsection{General theory}

Exceptional points are known to exist in a close vicinity of critical points and this is usually demonstrated, analytically or numerically, by letting some real parameters to be complex-valued \cite{triantafyllou1991,ghienne2020}. It is tempting to conduct a similar analysis with real-valued parameters, though the analysis is generally less trivial.
Progress can be made here by assuming that $\omega_{01}=\omega_{02}=\omega_{c}$ and $\kappa_1 \approx \kappa_2$ which means that  both pendulums are nearly identical, and by treating the normalized coupling strength,
\begin{equation}\label{eq:eps}
\epsilon = \frac{\kappa_1+\kappa_2}{2 \omega^2_{\mathrm{c}}},
\end{equation}
as a small parameter.
Once normalized, the eigenvalue problem takes the form (coupling coefficient $c_{12}$ is ignored here)
\begin{equation}
    \begin{bmatrix}
       \Lambda^2 + 2 \zeta_1 \Lambda 
       +1+\epsilon & - \epsilon\\
        - \epsilon & \Lambda^2 + 2 \zeta_2 \Lambda 
       +1+\epsilon
    \end{bmatrix}
        \mathbf{v} = \mathbf{0},
\end{equation}
where 
\begin{equation} \label{zetai}
\zeta_i = \frac{c_i}{2 \omega_c (m_i L_i^2 + I_i)},
\end{equation}
and we put $\Lambda = \lambda/\omega_{\mathrm{c}}$.
Following  standard perturbation theory, we look for  the first order approximation
\begin{equation} \label{lambda}
 \Lambda = \i + \alpha \epsilon + \mathcal{O}(\epsilon^2), \qquad \zeta_i=\alpha_i \epsilon + \mathcal{O}(\epsilon^2),
\end{equation}
and straightforward calculation yields
\begin{equation} \label{alpha}
 \alpha = \frac{\i -(\alpha_1 + \alpha_2) \pm \sqrt{(\alpha_1-\alpha_2)^2-1}}{2}.
\end{equation}
It follows that the perturbed eigenvalue remains a double root (up to first order) whenever $\alpha_i=\alpha^0_i$ satisfying
\begin{equation}\label{eq:zeta_eps}
|\alpha^0_1 - \alpha^0_2| = 1.
\end{equation}
This result shows there must be a good balance between the coupling strength and losses, i.e. $\zeta_i$ and $\epsilon$ must be of comparable amplitude.
The eigenvalue corresponding to the EP is given explicitly by
\begin{equation}
\lambda_0=-\sigma_0 + \i \omega_0 =-\omega_{\mathrm{c}}(\alpha^0_1 + \alpha^0_2 )\epsilon/2 
 + \i \omega_{\mathrm{c}}(1+ \epsilon/2).
\end{equation}
One can take advantage of the closed-form expression~\eqref{alpha} and consider a small perturbation  around the EP with respect to the two real-valued parameters $\alpha_1$ and $\alpha_2$. Straightforward calculation yields the Puiseux series (only the leading term is retained):
\begin{equation}
\lambda_i \approx \lambda_0 + (-1)^i \sqrt{(\alpha_1 -\alpha^0_1) - (\alpha_2 -\alpha^0_2) } \, \omega_c \epsilon /\sqrt{2}.
\end{equation}
Note that the term in the square root is purely real and can be canceled by choosing $\alpha_1 -\alpha^0_1 = \alpha_2 -\alpha^0_2$.
This is because points belonging to the continuous curves $\alpha_2=\alpha_1\pm 1$ are \emph{all} EPs.
This scenario is rather exceptional, and EPs are generally expected to  define a discrete set in the parameter space.

\subsection{Application to our experimental configuration}

These theoretical considerations seem restricted  to the ideal scenario of two weakly coupled identical pendulums. Nevertheless, they  provide a reliable description of the vibrational modes in our experiments, as shown in Sec.~\ref{sec:results}. By neglecting the damping coefficient of the first pendulum, i.e. $\alpha_1=\zeta_1=0$, one finds that an EP is found if $\alpha_2^0=1$ and consequently (see \eqref{zetai})
\begin{equation}
c_0 \approx 2 (m_2 L_2^2 +I_2) \omega_{\mathrm{c}} \epsilon,
\end{equation}
 where it is reminded that the length of the second pendulum must satisfy the condition that $\omega_{01}=\omega_{02}=\omega_{c}$ which holds if  $L_2=L_{\mathrm{c}}$. It follows that the coefficient $\alpha_2$ as defined in \eqref{lambda} can be simply expressed as the ratio 
 \begin{equation}
\alpha_2 = \frac{c_2}{c_0}.
\end{equation}
Finally, the trajectories of the two eigenvalues with respect to $c_2$ are given by 
\begin{equation} 
\lambda_i \approx \i \omega_0 -\sigma_0 \left (\alpha_2 + (-1)^i \sqrt{\alpha_2^2-1} \right ). 
\end{equation}

\section{A general method in order to find exceptional points with two real-valued parameters}
\label{app:real-valued_exceptional_point}

An EP corresponds to the solution of the two equations
\begin{equation}\label{eq:polsys}
    \mathbf{f} (\lambda; \mu, \nu) =
    \begin{bmatrix}
        p(\lambda; \mu, \nu) \\
        \partial_\lambda p (\lambda; \mu, \nu)
    \end{bmatrix} = \mathbf{0},
\end{equation}
where $\lambda$ is the eigenvalue which is generally complex-valued and $\mu$ and $\nu$ are two real parameters. We may assume that the system~\eqref{eq:polsys} is polynomial with respect to the parameters (which is the case here). If we imagine for the moment that only a single parameter is available, then Bézout theorem states that discrete solutions exist if the latter is allowed to be complex-valued. 
When two complex-valued parameters are available, solutions belong to a one-dimensional continuous domain. The difficulty arises when both parameters are restricted to be real-valued. A simple method is to define the variable  $\mathbf{x} (\nu) = [\lambda(\nu), \mu(\nu)]^{\mathrm{T}}$ which  implicitly depends on the free parameter $\nu$. The first step is to consider an initial guess $\nu_{\mathrm{ig}}$ and find a solution
$\mathbf{x}_{\mathrm{ig}}=\mathbf{x} (\nu_{\mathrm{ig}})$.
The second step is to track the solution path, while keeping $\nu$ real-valued, until a real-valued solution $\mu(\nu)$ is found.
Instead of solving the non-linear system~\eqref{eq:polsys} successively, the problem is recast into an ODE and solved by integration starting from the initial solution $\mathbf{x}_{\mathrm{ig}}$.
Along the solution path, we must have
\begin{equation}
        \frac{\dif}{\dif \nu} \mathbf{f}(\mathbf{x} (\nu), \nu)  = \mathbf{0}.
\end{equation}
Applying the multivariate chain rule yields the first-order ODE for the solution $\mathbf{x}$:
\begin{equation} \label{eq:numerical_continuation}
    \frac{\dif \mathbf{x}}{\dif \nu} = - \mathbf{J}_{\mathbf{f}}^{-1} (\mathbf{x}, \nu) \frac{\partial \mathbf{f}}{\partial \nu} (\mathbf{x}, \nu),
\end{equation}
where $\mathbf{J}_{\mathbf{f}} = \frac{\partial \mathbf{f}}{\partial \mathbf{x}}$ is the Jacobian matrix.
All derivatives involved are explicitly computed from Eq.~\eqref{eq:system_matrices}.
This evolution problem is solved numerically with the Runge-Kutta method.
Here, \texttt{pypolsys} python package\footnote{\url{https://github.com/nennigb/pypolsys}} is used to solve the multivariate polynomial system with the homotopy method. It is noteworthy that the characteristic polynomial may become ill-conditioned in the vicinity of an EP when the number of degrees of freedom is too large. In this latter case, the main steps of the method can be followed by restricting the search algorithm to a subset of eigenvalues \cite{ghienne2020, mailybaev2006}.

\begin{figure}[ht]
    \centering
    \includegraphics{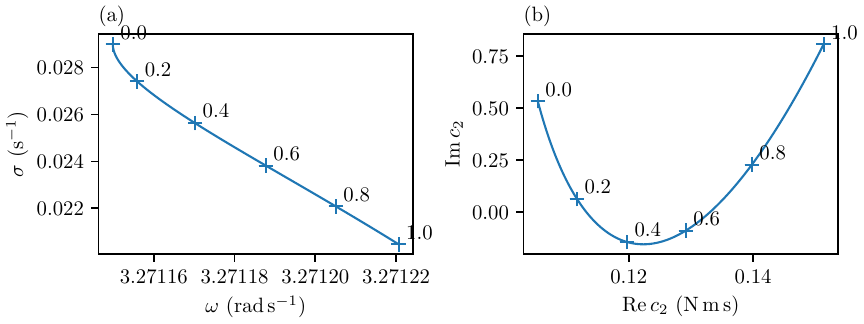}
    \caption{Complex solutions of the initial value problem from Eq.~\eqref{eq:numerical_continuation} where each marker indicates the value of $L_2$.}
    \label{fig:numerical_continuation}
\end{figure}

In order to illustrate this, the algorithm is applied to our configuration with $L_2=\nu$ and $c_2=\mu$ and results are shown in Fig.~\ref{fig:numerical_continuation}. All points of these curves are EP. The two EPs of interest correspond to a real-valued damping, $\mathrm{Im}(c_2)=0$. For the interested reader, the code is available at \url{https://github.com/nicolase7en/real-valued-ep2}.

\end{document}

%% file: tikz/coupled_pendulums.tex
\begin{circuitikz}
    \draw (-0.5, 0) -- (2.5, 0);
    \pattern[pattern=north east lines] (-0.5, 0) rectangle (2.5, 0.25);
    
    \path (0, 0) ++ (-105:4.5) node (m1) {} ;
    \draw (0, 0) -- (m1.center) ;
    \draw [black, fill=gray!50] (m1) circle (0.5) ;
    \draw (m1) node {$m_1$} ;
    \draw [dashed] (0, 0) -- (0, -4.5) node [midway, right]{$L_1$} ;
    \draw [->] (0, -3.5) arc (-90:-105:3.5);
    \path (0, 0) ++ (-97.5:3) node {$\theta_{1}$};
    
    \path (2, 0) ++ (-95:2) node (m2) {} ;
    \path (2, 0) ++ (-95:5) node (x) {} ;
    \draw (2, 0) -- (x.center) ;
    \draw [dashed] (2, 0) -- (2, -2) node [midway, right]{$L_2$} ;
    \draw [dashed] (2, -2) -- (2, -4.75) ;
    \draw [black, fill=gray!50] (m2) circle (0.5) ;
    \draw (m2) node {$m_2$} ;
    \draw [->] (2, -3.5) arc (-90:-95:3.5);
    \path (2, 0) ++ (-92.5:3) node {$\theta_{2}$};
    
    \draw [black, fill=gray!50] (1.25, 0) ++ (-95:4) -- ++ (-95:1.5) -- ++ (-5:1.5) -- ++ (85:1.5) -- ++ (175:1.5);
    
    \path (0, 0) ++ (-105:1) node (left) {} ;
    \path (2, 0) ++ (-95:1) node (right) {} ;
    \draw (left.center) to [spring=$k$] (right.center) ;
    \path (0, 0) -- (left.center) node [midway, left]{$d$} ;
    \path (2, 0) -- (right.center) node [midway, left]{$d$} ;
    
    \draw [black, fill=black] (2, -5.25) circle (0.05);
    \draw [black, thick] (2, -5.25) circle (0.15) node [above left] {$\mathbf{B}$};
\end{circuitikz}